\documentclass[12pt]{article}
\usepackage{epsfig,amssymb,euscript,bbold}
\usepackage{amsmath}
\newcommand{\half}{\tfrac{1}{2}}
\renewcommand{\d}{\partial}

\def\be#1\ee{\begin{equation}#1\end{equation}}
\newcommand{\bea}{\begin{eqnarray}}
\newcommand{\eea}{\end{eqnarray}}
\newcommand{\ba}{\begin{array}}
\newcommand{\p}[1]{(\ref{#1})}
\newcommand{\ea}{\end{array}}

\def\bbox{{\,\lower0.9pt\vbox{\hrule \hbox{\vrule height 0.2 cm
\hskip 0.2 cm \vrule height 0.2 cm}\hrule}\,}}
\newcommand{\dsl}{\pa \kern-0.5em /}

\newcommand{\nn}{\nonumber \\}
\newcommand{\zT}{\xi}

\newcommand{\tr}{{\rm tr}\,}

\def\a{{\alpha}}
\def\b{{\beta}}
\def\fuss{{\theta}}
\def\l{\lambda}

\def\ds{\raise.15ex\hbox{/}\kern-.57em\partial}
\def\Ds{\,\raise.15ex\hbox{/}\mkern-13.5mu D}
\font\mybb=msbm10 at 10pt
\def\bb#1{\hbox{\mybb#1}}

\def\bE {\bb{E}}

\begin{document}


\baselineskip 18pt


\begin{titlepage}
\vfill
\begin{flushright}
QMUL-PH-02-06\\
hep-th/0203255\\
\end{flushright}

\vfill

\begin{center}
\baselineskip=16pt
{\Large\bf pp-waves in 11-dimensions\\
\Large\bf with extra supersymmetry\\}
\vskip 10.mm
{Jerome P. Gauntlett$^{1}$ and Christopher M. Hull$^{2}$}\\
\vskip 1cm
{\small\it 
Department of Physics\\
Queen Mary, University of London\\
Mile End Rd, London E1 4NS, U.K.\\}
\vskip 0.5cm
and
\vskip 0.5cm
{\small\it Isaac Newton Institute for Mathematical Sciences\\
University of Cambridge\\
20 Clarkson Road, Cambridge, CB3 0EH, U.K.\\}
\vspace{6pt}
\end{center}
\vfill
\par
\begin{center}
{\bf ABSTRACT}
\end{center}
\begin{quote}
The Killing spinor equations for 
pp-wave solutions of eleven dimensional supergravity are analysed and
it is shown that there are solutions
that preserve 18, 20, 22 and 24 supersymmetries, in addition to the
generic solution preserving 16 supersymmetries and 
the Kowalski-Glikman solution preserving 32 supersymmetries.
\vfill
\vskip 5mm
\hrule width 5.cm
\vskip 5mm
{\small
\noindent $^1$ E-mail: j.p.gauntlett@qmul.ac.uk \\
\noindent $^2$ E-mail: c.m.hull@qmul.ac.uk \\
}
\end{quote}
\end{titlepage}
\setcounter{equation}{0}

\section{Introduction}

Eleven-dimensional supergravity has pp-wave solutions
\cite{Mwave}  
\begin{equation}
  \label{eq:metric}
  \begin{aligned}
  ds^2 &= 2 dx^+ dx^- + H(x^i,x^-) (dx^-)^2 + \sum_{i=1}^9 (dx^i)^2\\
  F_4&= dx^-\wedge \xi (x^i,x^-)
 \end{aligned}
\end{equation}
where
$H(x^i,x^-)$ obeys
\begin{equation}
  \label{eq:poisson}
  \bigtriangleup H= -\frac {1}{4} \|\xi \|^2 .
\end{equation}
Here $\bigtriangleup$ is the laplacian in the transverse euclidean
space $\bE^9$ with coordinates $x^i$ and $\xi (x^i,x^-)$ is (for
each $x^-$) a closed and coclosed $3$-form in $\bE^9$.  
This solution has at least 16 Killing spinors.

An interesting subclass of these metrics are those for which
\begin{equation}
  \label{eq:cwtype}
  H(x^i,x^-)= \sum_{i,j} A_{ij} x^i x^j
\end{equation}
where $A_{ij} = A_{ji}$ is a constant symmetric matrix \cite{FOPflux}. 
In particular, this class contains a maximally supersymmetric solution
with
  32 Killing spinors \cite{KG}  
\begin{equation}
  \label{eq:Aij}
  \begin{aligned}[m]
    A_{ij}& =
    \begin{cases}
      -\tfrac19 \mu^2 \delta_{ij} & i,j=1,2,3\\
      -\tfrac1{36} \mu^2 \delta_{ij} & i,j=4,5,\dots,9
    \end{cases}\\
    \xi &=\mu dx^1\wedge dx^2\wedge dx^3~,
  \end{aligned}
\end{equation}
where $\mu$ is a parameter which can be set to any nonzero value by a
change of coordinates. 

In \cite{IIB}, a similar maximally supersymmetric solution of IIB
supergravity
was found, and in \cite{penlimit,mald} 
it was shown that both of these solutions arise as
Penrose limits \cite{pen} of maximally supersymmetric $AdS\times Sphere$
solutions. 
The IIB string theory in this  background can
be exactly solved  \cite{mets}  and   is dual to
a certain subsector of $N=4$ supersymmetric Yang-Mills theory
\cite{mald}.
Subsequent work developing these ideas includes
\cite{Blau:2002rg}-\cite{maldlatest}.

Given that pp-waves generically preserve at least half of the
supersymmetries, and for special cases preserve all of the
supersymmetries,
it is natural to ask whether there are similar solutions preserving
fractions $\nu$ of the supersymmetry with $1/2< \nu < 1$.
In \cite{extra}, it was argued that configurations preserving 
such fractions of supersymmetry could arise in M-theory.
In \cite{pope} it was shown that such configurations 
do indeed arise as IIB pp-waves, and a pp-wave of M-theory
preserving 3/4 supersymmetry was presented in \cite{michelson}.
The purpose here is to investigate pp-wave solutions
of 11-dimensional supergravity in more detail, and to show that 
the fractions 9/16,5/8,11/16 can also arise in addition to 3/4,1/2 and
1.

Our ansatz is 
\begin{equation}
  \label{eq:cwsoll}
  \begin{aligned}[m]
    ds^2 &= 2 dx^+ dx^- + \sum_{i,j} A_{ij} x^i x^j (dx^-)^2 +  \sum_i
    dx^i dx^i\\
    F &=  dx^- \wedge \zT,
  \end{aligned}
\end{equation}
where $\zT$ is a 3-form on $\bE^9$ with constant coefficients. This is
a supersymmetric solution of eleven-dimensional supergravity,  provided
that
\begin{equation}
  \label{eq:conn}
  \tr A = -\half \|\zT\|^2=- \tfrac1{12} \zT_{ijk}
  \zT^{ijk}~,
\end{equation}
with $i,j,k=1,\dots,9$.
We seek the conditions for such solutions to admit Killing spinors,
following the analysis of \cite{FOPflux} and \cite{IIB},\cite{pope}.
We will see that this occurs for specific choices of $\zT,A$.

The Killing spinors $\varepsilon$ satisfy the equation
\begin{equation}
  \label{eq:Killing}
  \nabla_M \varepsilon = \Omega_M \varepsilon~,
\end{equation}
where $\nabla$ is the spin connection and
\begin{equation}
  \label{eq:Omega}
  \Omega_M = \tfrac1{288}
  \left( \Gamma_M{}^{PQRS} - 8 \delta^P_M\Gamma^{QRS}\right)F_{PQRS}~,
\end{equation}
In the frame
\bea
e^+&=&dx^++\tfrac1{2}\sum_{i,j}A_{ij}x^ix^jdx^-\nn
e^-&=&dx^-\nn
e^i&=&dx^i
\eea
the only nonvanishing components of the spin
connection are
\begin{equation}
  \omega^{+i} = \sum_j A_{ij} x^j dx^-~.
\end{equation}
We also have
\begin{equation}
  \label{eq:Omegaexplicit}
  \begin{aligned}[m]
    \Omega_+ &= 0\\
    \Omega_- &= - \tfrac{1}{12} \Theta \left( \Gamma_+ \Gamma_- + 1
    \right) \\
    \Omega_i &= \tfrac{1}{24} (3\Theta \Gamma_i +  \Gamma_i
\Theta)\Gamma_+
  \end{aligned}
\end{equation}
where the indices on the left-hand-side are co-ordinate indices, while
the
indices on the gamma-matrices here, and for the rest of the paper, 
are frame indices, and
\begin{equation}
\Theta= \tfrac1{6} \zT_{ijk} \Gamma^{ijk}~.
\end{equation}

For any such solution, it is simple to see that
there are always 16 ``standard'' Killing spinors
satisfying 
\begin{equation}
  \label{eq:kilpro}
  \Gamma_+\varepsilon=0~.
\end{equation}
Explicitly they are given by 
\begin{equation}
\label{eq:hppspinora}
  \varepsilon = \exp\left(-\tfrac{1}{4} x^- \Theta\right)\psi~,
\end{equation}
for some constant spinor $\psi$ such that $\Gamma_+\psi=0$.

Next we look for ``extra''
Killing spinors with $\Gamma_+\varepsilon
\ne 0$. 
It was shown in  \cite{FOPflux} that {\it any} Killing spinor
is of the form
\begin{equation}
  \label{eq:kill0}
  \varepsilon = \left(1 + \sum_i x^i \Omega_i \right) \chi~,
\end{equation}
where the spinor $\chi$ only depends on $x^-$.
The dependence on $x^-$ is determined by
\begin{equation}
  \d_- \varepsilon = - \half \sum_{i,j} A_{ij} x^j \Gamma_+ \Gamma_i
  \varepsilon - \tfrac1{12} \left( \Gamma_+ \Gamma_- + 1 \right)
  \Theta \varepsilon~,
\end{equation}
which, using \eqref{eq:kill0},   gives 
\begin{multline}
  \label{eq:xks}
  \frac{d}{dx^-}\chi = - \tfrac1{12} \Theta \left(1 + \Gamma_+
    \Gamma_-\right) \, \chi\\
  + \sum_i x^i \left( - \half \sum_j A_{ij} \Gamma_+ \Gamma_j +
    \tfrac1{12}  \Omega_i \Theta - \tfrac14 \Theta \Omega_i\right)\,
  \chi~.
\end{multline}
As $\chi$ is independent of $x^i$, 
this can be decomposed into a piece independent of $x^i$, and a piece
that is linear in $x^i$. The piece independent of $x^i$ 
is
\begin{equation}
\label{eq:consp}
\frac{d}{dx^-}\chi = - \tfrac1{12}
\left( \Gamma_+ \Gamma_- + 1 \right) \Theta  \chi
\end{equation}
which determines the $x^-$ dependence of $\chi$. 
The part linear in $x^i$ gives
\begin{equation}
\label{eq:aeqn}
\left( - 144 \sum_j A_{ij}\Gamma_j +
    9\Theta^2\Gamma_i +6\Theta\Gamma_i\Theta +\Gamma_i\Theta^2\right)\,
  \Gamma_+\chi~=0.
\end{equation}
and we now proceed to analyse this.

We choose a representation of the $32\times 32$ Dirac
matrices $\Gamma_M$ in which
\begin{equation}
 \label{eq:abc}
 \Gamma _i = \gamma _i \otimes \sigma _3, \qquad
\Gamma _\pm  = 1 \otimes \sigma _\pm
\end{equation}
where
$i,j=1,...,9$, $\gamma_i$ are $16\times 16$ gamma matrices for
$SO(9)$, $(\sigma_1,\sigma_2,\sigma_3)$ are $2\times 2$ Pauli matrices,
with
$\sigma _\pm = \frac{1}{\sqrt 2}(\sigma _1 \pm i \sigma _2)$.
A 32-component spinor $\chi$ then decomposes into two $SO(9)$ spinors,
$\chi_\pm$:
\begin{equation}
\label{eq:fghdsa}
\chi = (\chi_+,\chi_-), \qquad \Gamma _-\chi = {\sqrt 2}
(0,\chi_+),\qquad 
\Gamma_+\chi =
 {\sqrt 2}(\chi_-,0)
\end{equation}
Then
\begin{equation}
\label{eq:po}
\Theta= \fuss\otimes \sigma _3, \qquad 
\fuss=\tfrac1{6} \zT_{ijk} \gamma^{ijk}
\end{equation}
and $\fuss_{\a\b} $ is an antisymmetric $16\times 16$ matrix, where
$\a,\b=1,...,16$ are $SO(9)$ spinor indices.
Equation \p{eq:consp} implies
\begin{equation}
\label{eq:hppspinorab}
  \chi _+ = \exp\left(-\tfrac{1}{4} x^- \theta\right)\psi_+~,
\qquad
  \chi _- = \exp\left(-\tfrac{1}{12} x^- \theta\right)\psi_-~,
\end{equation}
for constant 16-component  spinors $\psi _\pm$.
Equation \p{eq:aeqn} imposes no further conditions on $\chi_+$.

A Killing spinor in 11 dimensions 
will only give rise to a Killing spinor in the theory obtained by 
  dimensional reduction in a direction generated by a Killing vector
  if the Lie derivative in the Killing direction
  of that spinor vanishes.
The spinors $\psi_\pm$ that are anihilated by $\theta$
give Killing spinors that are independent of 
$x^-$ and hence survive under dimensional reduction in the $x^-$
direction. The standard Killing spinors, parametrised by $\psi_+$,
are independent of $x^i$ while the $x^i$ dependence of any extra 
Killing spinors, parametrised by $\psi_-$, are encoded in \p{eq:kill0}.
If the matrix $A$ is such that $\partial_i$ is a Killing vector then
the Killing spinors independent of $x^i$ will survive under 
dimensional reduction on this Killing vector.


Next, we need to specify our ansatz for $\zT$ and hence $\Theta$.
Any anti-symmetric matrix $L_{\a\b}$ 
can be written in terms of a 2-form $L_{ij}$ and a 3-form $L_{ijk}$
as
\begin{equation}
\label{eq:ells}
L_{\a\b}= \frac {1}{2} 
L_{ij} (\gamma ^{ij})_{\a\b}+\frac {1}{6}  L_{ijk} (\gamma
^{ijk})_{\a\b}
 \end{equation}
This gives a decomposition of 
the Lie algebra of $SO(16)$ (the $16\times 16 $ antisymmetric matrices
 $L_{\a\b}$) into the 
maximal $Spin(9)$ subalgebra (the $9\times 9 $ antisymmetric matrices
 $L_{ij}$), and its complement (specified by the 3-forms
$L_{ijk}$).
occured in
Now $SO(16)$ has rank 8 while $Spin(9)$ has rank 4, so 
any Cartan subalgebra of $SO(16)$ is
generated, for some $n\le 4$, by $n$ commuting generators of
$Spin(9)$ corresponding to  $n$ 2-forms, and $8-n$
commuting elements from the complement of $Spin(9)$, corresponding to
$8-n$ 3-forms.
Only the cases $n=4$ and $n=1$ occur.
A convenient choice for the Cartan subalgebra with $n=4$
is the commuting  set of four generators
\begin{equation}
 \label{eq:comm}
 (\gamma ^{ 12})_{\a\b} ,\qquad (\gamma ^{34 })_{\a\b} 
,\qquad (\gamma ^{56 })_{\a\b} 
,\qquad (\gamma ^{78 })_{\a\b} 
\end{equation}
of $Spin(9)$,
together with
\begin{equation}
 \label{eq:commtwo}
 (\gamma ^{ 129})_{\a\b} ,\qquad (\gamma ^{34 9})_{\a\b} 
,\qquad (\gamma ^{56 9})_{\a\b} 
,\qquad (\gamma ^{789 })_{\a\b} 
\end{equation}
A convenient basis with $n=1$
consists of the
$Spin(9)$ generator $(\gamma ^{89 })_{\a\b} $
together with the seven 3-forms
\begin{equation}
\label{eq:commtwoe}
 \begin{aligned}[m]
 & 
  (\gamma _{ 123})_{\a\b},\qquad 
  (\gamma _{ 145})_{\a\b}, \qquad 
  (\gamma _{ 167})_{\a\b}, \qquad 
  (\gamma _{ 246})_{\a\b}
\\
&
  (\gamma _{ 257})_{\a\b}, \qquad 
  (\gamma _{ 347})_{\a\b}, \qquad 
  (\gamma _{ 356})_{\a\b}\\
  \end{aligned}\end{equation}

A basis in spin-space can be chosen to bring any given
anti-symmetric $\fuss_{\a\b}$ to skew-diagonal form
with skew eigenvalues $\lambda_1,...,\lambda_8$, $\fuss= \epsilon
\otimes \Lambda$, where
$ \epsilon=i \sigma _2$ and  $\Lambda$
is the $8\times 8$ diagonal matrix
$\Lambda=diag(\lambda_1,...,\lambda_8)$.
 Then the eigenvalues of the
symmetric matrix
$\fuss^2$ are
$-\lambda_I^2$,
$I=1,...,8$, each with degeneracy 2.
The set of such skew-diagonal matrices generate an 8-dimensional Cartan
subalgebra of $SO(16)$, and so can be decomposed into
$n$ 2-forms and $8-n$ 3-forms where either $n=1$ or $n=4$, and the set
of 
$\fuss_{\a\b}$ are stratified into distinct orbits with  
$n=1$ or $n=4$.

Now the flux $\zT_{ijk}$ determines
a $16\times 16 $ anti-symmetric matrix
$\fuss_{\a\b}$ from \p{eq:po}.
However, it is not an arbitrary antisymmetric matrix, but one for which
the 2-form part vanishes, i.e. one satisfying the constraint
$(\gamma ^{ij})^{\a\b}\fuss_{\a\b}=0$.
If it occurs in the $n=4$ orbit, 
it can be written in terms of 
four linearly independent 3-forms
using \p{eq:ells}, and using $SO(9)$ transformations, these
can be arranged to be precisely the
generators in  \p{eq:commtwo}. 
That is, one can choose bases for spin-space and the
tangent space such  that the generators 
\p{eq:comm},\p{eq:commtwo} are skew-diagonal and $\fuss_{\a\b}$
is a linear combination of the
3-form generators \p{eq:commtwo} alone, so that there are constants
$m_1,m_2,m_3,m_4 $ such that
\begin{equation}
 \label{eq:wis}
 \fuss_{\a\b}=m _1
  (\gamma _{ 129})^{ab}+m _2 (\gamma _{34 9})^{ab} 
+m _3 (\gamma _{56 9})^{ab} 
+m _4 (\gamma _{789 })^{ab} 
\end{equation}
The skew eigenvalues $\lambda_1,...,\lambda _8$ of $\fuss$ are then,
in a convenient basis, given by
\begin{equation}
  \label{eq:wisss}
  \begin{aligned}[m]
 \lambda_1&=-m_1-m_2+m_3-m_4\\
 \lambda_2&=m_1+m_2-m_3-m_4\\
 \lambda_3&=m_1+m_2+m_3-m_4\\
 \lambda_4&=-m_1-m_2-m_3-m_4\\
 \lambda_5&=-m_1+m_2+m_3+m_4\\
 \lambda_6&=m_1-m_2-m_3+m_4\\
 \lambda_7&=m_1-m_2+m_3+m_4\\
 \lambda_8&=-m_1+m_2-m_3+m_4\\
  \end{aligned}
\end{equation}

Similarly, if $\fuss_{\a\b}$ lies in the $n=1$ orbit, 
one can choose bases such that
\begin{equation}
\label{eq:wist}
 \begin{aligned}[m]
 \fuss_{\a\b}&=n _1
  (\gamma _{ 123})_{\a\b}+
n _2
  (\gamma _{ 145})_{\a\b}+
n _3
  (\gamma _{ 167})_{\a\b}+
n _4
  (\gamma _{ 246})_{\a\b}
\\
&+
n _5
  (\gamma _{ 257})_{\a\b}+
n _6
  (\gamma _{ 347})_{\a\b}+
n _7
  (\gamma _{ 356})_{\a\b}\\
  \end{aligned}\end{equation}
for some
constants
$n_1,n_2,...,n_7$.
The skew eigenvalues $\lambda_1,...,\lambda _8$ of $\fuss$ are then,
in a convenient basis, given by
\begin{equation}
  \label{eq:wissst}
  \begin{aligned}[m]
 \lambda_1&=-n_1-n_2-n_3-n_4+n_5+n_6+n_7\\
 \lambda_2&=-n_1+n_2+n_3+n_4-n_5+n_6+n_7\\
 \lambda_3&=n_1+n_2-n_3-n_4-n_5-n_6+n_7\\
 \lambda_4&=n_1-n_2+n_3+n_4+n_5-n_6+n_7\\
 \lambda_5&=n_1-n_2+n_3-n_4-n_5+n_6-n_7\\
 \lambda_6&=n_1+n_2-n_3+n_4+n_5+n_6-n_7\\
 \lambda_7&=-n_1+n_2+n_3-n_4+n_5-n_6-n_7\\
 \lambda_8&=-n_1-n_2-n_3+n_4-n_5-n_6-n_7\\
  \end{aligned}
\end{equation}

The upshot of this analysis is that 
without loss of generality we can take the flux to be such that
$\fuss$ is given either by \eqref{eq:wis}
in terms of four coefficients $m _a$, corresponding to
\begin{equation}
  \label{eq:thisss}
  \zT = m_1 dx^{129} +m_2 
 dx^{349}  +
m_3 dx^{569}   
+m_4 dx^{789} 
\end{equation}
or by
\eqref{eq:wist}
in terms of seven coefficients $n _a$, corresponding to
\begin{equation}
  \label{eq:thistle}
  \begin{aligned}[m]
 \zT &= n_1 dx^{123} +n_2 
 dx^{145}  +
n_3 dx^{167}   
+n_4 dx^{246} \\
&
+
 +n_5
 dx^{257}  +
n_6 dx^{347}   
+n_7 dx^{356}
 \\
  \end{aligned}
\end{equation}
Here $dx^{ijk}=dx^i\wedge dx^j\wedge dx^j$.
It is now straightforward to analyse the supersymmetry of the solution
given by the ansatz \p{eq:thisss} or \p{eq:thistle}.

It will be useful to define 
$\theta _{(i)}$
by
\begin{equation}
\label{eq:fgfdd}
\theta \gamma _i =\gamma _i \theta _{(i)}
 \end{equation}
for each $i$, so that
\begin{equation}
\label{eq:hjjgd}
 9\Theta^2\Gamma_i +6\Theta\Gamma_i\Theta +\Gamma_i\Theta^2
=\gamma _i( 9\theta _{(i)}^2 +6\theta _{(i)}\theta +\theta ^2)\otimes 1
\end{equation}
If $\theta$  is chosen so that $\theta _{(i)}$ commutes with $\theta$,
as can be shown to be   the case for the ans\" atze for $\theta$ 
\p{eq:thisss} or \p{eq:thistle},
this can be rewritten as
\begin{equation}
\label{eq:dasds}
\gamma _i U_{(i)}^2\otimes 1, \qquad U_{(i)}\equiv  3\theta _{(i)} 
+\theta  
\end{equation}
Then \eqref{eq:aeqn}
implies 
\begin{equation}
\label{eq:aeqna}
\left( - 144 \sum_j A_{ij}\gamma_j + \gamma _i U_{(i)}^2
     \right)\,
 \chi_-~=0.
\end{equation}
for each $i=1,...,9$, with, as seen above, no condition on $\chi_+$.
In a basis in which $U_{(i)}^2 $ is diagonal for each $i=1,...,9$,
this can only have solutions if
$A_{ij}$ is also diagonal,
\begin{equation}
\label{eq:ais}
A_{ij} = - diag ( \mu _1^2, \mu _2 ^2,..., \mu _9^2)
 \end{equation}
for some constants $\mu_i$. 
Let the skew eigenvalues of  $U_{(i)} $ be $\rho_{I(i)}$, $I=1,...,8$,
so that $U_{(i)} ^2$ is a symmetric matrix with   eigenvalues
$-\rho_{I(i)}^2$, each with 2-fold degeneracy.
Then 
if $\chi_-$ is chosen as an eigenvector
$\chi_I$ satisfying
\begin{equation}
\label{eq:sdjf}
U_{(i)}^2\chi_I=-\rho_{I(i)}^2\chi_I
\end{equation}
for some $I$, then it defines a Killing spinor providing that
$A_{ij}$ is given by \p{eq:ais} with
the 9 coefficients $\mu _i$ determined to be
\begin{equation}
\label{eq:muis}
144 \mu _i^2= \rho_{I(i)}^2
\end{equation}
There will be (at least) 2 such extra Killing spinors,
as each eigenvalue has (at least) two-fold degeneracy. 

Given this choice of $A_{ij}$, a second pair of eigenspinors
$\chi_J$ ($J\ne I$) will also give  extra Killing spinors if and only if
\begin{equation}
\label{eq:rhoeq}
\rho_{J(i)}^2=\rho_{I(i)}^2
\end{equation}
for each $i$.
If there is an $N$-fold degeneracy in the eigenvalues $\rho_{I(i)}^2$, 
\begin{equation}
\label{eq:muisagain}
144 \mu _i^2= \rho_{J_1(i)}^2= \rho_{J_2(i)}^2=...= \rho_{J_N(i)}^2
\end{equation}
for all $i$, then
there are $2N$ such extra Killing spinors, and 
the solution will have a total of $16+2N$ Killing spinors.

 Given a flux defined by $\zT$, and any choice of $I$, the
matrix
$A_{ij}$ can be chosen as in \p{eq:ais},\p{eq:muis} so that the two
spinors
$\chi_I$  
with  $\Theta^2$ eigenvalues $-\lambda _I ^2$
are Killing spinors.
 Next we turn to the conditions for degeneracy, \p{eq:rhoeq}.
In a basis in which the anti-symmetric matrix $\theta$ is skew-diagonal
with skew eigenvalues
$\lambda _I$, then  $\theta_{(i)}$
is also skew-diagonal for either ansatz \p{eq:thisss} or \p{eq:thistle}; 
let its eigenvalues
be $\lambda_{I(i)}$, and define $k_{I(i)}$ by
\begin{equation}
\lambda _I+\lambda_{I(i)}=2 k_{I(i)}~.
\end{equation}
Then
\begin{equation}
\label{eq:fgldf}
\rho_{I(i)}= 3\lambda_{I(i)}+\lambda _I = -2(\lambda _I-3 k_{I(i)})~.
\end{equation}
For ansatz \p{eq:thisss} $\rho _{I(9)}=4\lambda _I$, while for ansatz
\p{eq:thistle},
$\rho _{I(8)}=\rho _{I(9)}=-2\lambda _I$, so that
\p{eq:rhoeq} implies
\begin{equation}
\label{eq:gfjsd}
 \lambda _I^2=\lambda _J^2
\end{equation}
and hence either $\lambda _I= \lambda _J$, or $\lambda _I= -\lambda _J$.
If $\lambda _I= \lambda _J=0$, then 
\p{eq:rhoeq},\p{eq:fgldf} implies 
\begin{equation}
\label{eq:asdagf}
k_{I(i)}^2=k_{J(i)}^2
\end{equation}
for each $i$.
If $\lambda _I= \lambda _J\ne0$, then \p{eq:rhoeq},\p{eq:fgldf} implies
\begin{equation}
\label{eq:fgldftwo}
(k_{I(i)}-k_{J(i)})(3k_{I(i)}+3k_{J(i)}-2\lambda _I)=0
\end{equation}
so that either
$k_{I(i)}=k_{J(i)}$, or $3k_{I(i)}+3k_{J(i)}=2\lambda _I$.
Similarly, if
$\lambda _I= -\lambda _J\ne 0$, then
 either
$k_{I(i)}=-k_{J(i)}$, or $3k_{I(i)}-3k_{J(i)}=2\lambda _I$.

Let us now analyse the 4-parameter ansatz \p{eq:thisss} in detail,
before briefly returning to the 7-parameter case \p{eq:wist} at the end.
For \p{eq:thisss}, the 
eigenvalues
$\lambda _I$ \p{eq:wisss} are
of the form
\begin{equation}
\label{eq:fgldfthree}
\lambda _I= \sum _a L_{Ia}m_a
\end{equation}
where
each coefficient $L_{Ia}=\pm 1$.
Then for $i=1,...,8$,
\begin{equation}
\label{eq:fgldffour}
k_{I(i)}=L_{Ia}m_a, \qquad {\rm with}\qquad a=[(i+1)/2]
\end{equation}
where $a$ is the integer part of $(i+1)/2$, so that for example
$k_{I(3)}=k_{I(4)}=L_{I2}m_2$.
Then $k_{I(i)}^2=m_a^2$ with $a=[(i+1)/2]$.
This implies
\eqref{eq:asdagf} for all $I,J$.
Thus if $\lambda _I= \lambda _J=0$, then 
 $\chi_I $ and $\chi_J$ are both Killing spinors, provided the $\mu_i$
are chosen as in \p{eq:muis}, and $N\ge 2$. 

For at least one value of $a$, $L_{Ia}=L_{Ja}$, while for at least one
other value of $a$
$L_{Ia}=-L_{Ja}$, so that for some
$i$, $k_{I(i)}=k_{J(i)}$ and for others
$k_{I(i)}=-k_{J(i)}$.
Consider next the case
$\lambda _I= \lambda _J\ne 0$.
Then for those $i$ for which 
$k_{I(i)}=k_{J(i)}$, \p{eq:rhoeq} is satisfied, while for 
those $i$ for which 
$k_{I(i)}=-k_{J(i)}$, \p{eq:fgldftwo} implies that
$k_{I(i)}=0$, which in turn implies $m_a=0$ for
those $a=[(i+1)/2]$.
Similarly, if
$\lambda _I= -\lambda _J\ne 0$, then for those $i$  such that
$k_{I(i)}=k_{J(i)}$ the masses $m_a$ must vanish for
those $a=[(i+1)/2]$.

Thus  if $A_{ij}$ is chosen so that
$\chi_-=\chi _I$ defines an extra Killing spinor, 
then the conditions (with the 4-parameter ansatz) that $\chi _J$ gives
rise to a further   Killing spinor
are as follows.
First, 
$\lambda _I^2=\l_J^2$.
If $\lambda _I=0$, no further conditions are needed.
If $\lambda _I\ne 0$, then one or more of the mass parameters $m_a$ must
be set to zero.

We will now summarise the degeneracy numbers $N$ for each case, and
so give the total number $16+2N$ of Killing spinors.
First, for generic
$A_{ij}$, there are no extra Killing spinors,
$N=0$, and there are just the $16$ standard  Killing spinors. If the
$A_{ij}$ are chosen as in \p{eq:ais},\p{eq:muis} 
for some $I$, then $N\ge 1$
and there are at least two extra Killing spinors, giving at least 18
supersymmetries. If one of the   parameters is set to zero, $m_4=0$
say, then
$N=2$ and there are at least  20   Killing spinors.
If two   parameters are set to zero, $m_3=m_4=0$
say, then $N=4$ and there are 24 supersymmetries.
If three masses are set to zero, then $N=8$ and the maximally
supersymmetric solution of \cite{KG} is recovered.
%
If, on the other hand, the parameters  are chosen to give 
$\lambda _I=0$ but with all $m_a$ non-zero, then there are no extra
supersymmetries.
If the masses are chosen so that in addition some of the other
eigenvalues
$\lambda _J $ vanish, then there will be further supersymmetries.
The parameters $m_a$ can be chosen with them all non-zero so that 1,2 or
3
of the eigenvalues
$\lambda _I$ vanish, giving $N=1,2$ or $3$ and so
18,20 or 22 supersymmetries respectively.

The Killing spinors take the   form \eqref{eq:kill0}
where the spinor $\chi= (\chi_+,\chi_-)$ is given in terms of constant
spinors $\psi_+,\psi_-$ by
\eqref{eq:hppspinorab}.
The spinors $\psi_+$ are unconstrained, giving 16 standard Killing
spinors, while
the spinors $\psi_-$ are restricted to be $2N$ eigenspinors of
$U_{(i)}^2$
with eigenvalue
\eqref{eq:muis}.
Then  
on this $2N$ dimensional space,
$U_{(i)} =12 \mu_{i}J$, where
$J=\epsilon\otimes 1_{N\times N}$
and satisfies $J^2=-1$.
Then $\varepsilon=(\varepsilon_+,\varepsilon_-)$ with
\begin{equation}
  \label{eq:killfd}
  \varepsilon _- = \chi_-, \qquad
 \varepsilon _+ = \chi_+ + \frac{1}{\sqrt 2}\sum_i x^i \gamma_i \mu _i
J
\chi_-
\end{equation}
Similarly, on this $2N$-dimensional space,
$\theta$ is skew-diagonal with $\theta =\lambda J$ for some $\lambda$,
so that 
 \begin{equation}
\label{eq:fdjlgsds}
\chi _- = \exp\left(-\tfrac{1}{12} x^- \theta\right)\psi_-~
= \cos (\lambda x^-/12)\psi_- -
\sin (\lambda x^-/12)J\psi_-
\end{equation}
and $\psi_-$ is restricted to lie in the  $2N$-dimensional eigenspace.
Note that if $\lambda=0$ then the $\chi_-$ are independent of 
$x^-$. Similarly, $\psi_+$ can be decomposed into eigenspinors of
$\theta$,
allowing the exponential in \eqref{eq:hppspinorab} to be calculated
explicitly.  

We now present explicit examples of the above cases.
Let us first consider the cases with more than 
18 supersymmetries obtained when one of the $m_a$ is zero for
generic non-vanishing $\lambda_I$.
Without loss of generality we set $m_4=0$. For this case we have 
$\lambda_1=-\lambda_2$ which corresponds to preservation of 20
supersymmetries in general. The solution is given by
\begin{equation}
  \label{eq:firstex}
  \begin{aligned}[m]
 \zT&=m_1 dx^{ 129}+m_2 dx^{34 9}+m _3 dx^{56 9}\\
   H=&-\tfrac{1}{36}(2m_1-m_2+m_3)^2(x_1^2+x_2^2)\\
    &-\tfrac{1}{36}(-m_1+2m_2+m_3)^2(x_3^2+x_4^2)\\
&-\tfrac{1}{36}(-m_1-m_2-2m_3)^2(x_5^2+x_6^2)\\
&-\tfrac{1}{36}(-m_1-m_2+m_3)^2(x_7^2+x_8^2+4x_9^2)\\
\end{aligned}
\end{equation}
Note that in general the extra Killing spinors will depend
on all coordinates. If we set $m_1=m_2=-m_3$ then they are independent 
of $x_1,\dots ,x_6$. On the other hand if we set $m_3=m_1+m_2$ (which
means $\lambda_1=\lambda_2=0$) then they are
independent of $x_7,x_8,x_9$ and $x^-$ and moreover four of the 16
Killing spinors \p{eq:hppspinora} are then also independent of $x^-$. 
This   has a metric which is the product of
a pp-wave metric in eight dimensions with $\bE^3$, but the flux has
non-trivial components
in one of the
directions in $\bE^3$, so that this solution can be regarded as a
product of 
a nine-dimensional solution with $\bE^2$.

If we further set $m_3=0$
then the solution preserves 24 supersymmetries in general and the
extra Killing spinors will still depend on all coordinates.
If $2m_1=m_2$ then the extra Killing spinors are independent of 
$x_1,x_2$. On the other hand if $m_1=-m_2$ (which means 
$\lambda_1=\lambda_2=\lambda_3=\lambda_4=0$) then they
are independent of $x_5,x_6,x_7,x_8,x_9$ and $x^-$. In this case
8 of the 16 standard Killing spinors \p{eq:hppspinora} are also
independent of $x^-$.
This has a metric which 
 is the product of
a pp-wave metric in six dimensions with $\bE^5$, but again the flux 
has non-trivial components
in one of the
directions in $\bE^5$.
Finally, if we further set either $m_2=0$ or $m_1=0$
we recover the maximally supersymmetric solution of \cite{KG}.

Next we consider cases of more than 18 supersymmetries with some
$\lambda_I=0$. In this case the extra Killing spinors defined by
$\chi_I$ do
not depend on $x^9$ or $x^-$. 
If we set $\lambda_1=\lambda_2=0$ then we deduce
$m_4=0$ which leads us back to one of the cases   above with 20
supersymmetries. 
On the other hand
if we set $\lambda_1=\lambda_3=0$, which can be achieved by setting
$m_1=-m_2$ and $m_3=m_4$, then we get something new. The solution now
takes the form
\begin{equation}
  \label{eq:secex}
  \begin{aligned}[m]
 \zT&=m_1 (dx^{ 129}- dx^{34 9})+m _3 (dx^{56 9}+dx^{789})\\
   H=&-\tfrac{1}{4}(m_1)^2(x_1^2+x_2^2 +x_3^2+x_4^2)\\
    &-\tfrac{1}{4}(m_3)^2(x_5^2+x_6^2+x_7^2+x_8^2)\\
\end{aligned}
\end{equation}
and generically preserves 20 supersymmetries. Note that for this case
4 of the 16 standard 
Killing spinors \p{eq:hppspinora} are also independent of $x^-$.

An interesting special
case is when in addition we have $\lambda_1=\lambda_3=\lambda_5=0$ 
which can be achieved by setting $m_1=-m_2=m_3=m_4$. The solution
then takes the form
\begin{equation}
  \label{eq:thirdtex}
  \begin{aligned}[m]
 \zT&=m_1 (dx^{129}- dx^{34 9}+dx^{56 9}+dx^{789})\\
   H=&-\tfrac{1}{4}(m_1)^2(x_1^2+x_2^2
+x_3^2+x_4^2+x_5^2+x_6^2+x_7^2+x_8^2)\\
\end{aligned}
\end{equation}
and preserves 22 supersymmetries. For this case the 
extra 6 Killing spinors are independent of $x^9$ and 
$x^-$ but are dependent on $x_1,\dots,x_8$. 
In addition 6 of the 16 standard Killing spinors 
\p{eq:hppspinora} are also independent of $x^-$.
Note that if we reduce on the $x_9$ direction
we obtain a type IIA pp-wave solution that preserves 22 supersymmetries.
This solution does not have a further Killing direction 
and so T-duality cannot be used to relate this to a IIB pp-wave
solution.
Thus such type IIA pp-wave solutions with 22 supersymmetries could not
be
obtained by starting with IIB solutions and T-dualising as in
\cite{michelson}.
Note that if we
dimensionally reduce the D=11 solution along the $x^-$ direction
we obtain a type IIA D0-brane solution that preserves 12
supersymmetries.

Finally we turn to the 7-parameter ansatz \p{eq:wist}.
For generic parameters, the $A_{ij}$ can again be chosen to give 
18 supersymmetries. Note that the expression for the Killing spinors
\p{eq:killfd},\p{eq:fdjlgsds}
discussed above for the 4-parameter case is also valid for this case.
In the 7-parameter case, the analysis of the conditions for further
supersymmetries
is more complicated, but some simple cases  can be analysed.
If certain sets of four of the parameters vanish, e.g. if
$n_4=n_5=n_6=n_7=0$, then the ansatz
is equivalent (after a relabelling) to the ansatz \p{eq:wisss} with
one of the parameters vanishing $m_a$, and hence gives
at least 20 supersymmetries as discussed above.
On the other hand, if certain sets of three parameters vanish
e.g. $n_1=n_2=n_3=0$, then there are different configurations with 
20 supersymmetries.
The special case
in which $n_4=n_5=-n_6=n_7=n$ leads to a solution
\begin{equation}
  \begin{aligned}[m]
 \zT &= n(dx^{246}+ dx^{257}  -dx^{347} +dx^{356})\\
   H=&-n^2(x_2^2 +x_3^2)\\
\end{aligned}
\end{equation}
In this case $\lambda_6=\lambda_3=0$ and $\chi_6$ and $\chi_3$ are the
extra 
Killing spinors, which
are independent of $x^-$. Since
$\lambda_1=\lambda_2=\lambda_7=\lambda_8=0$
we conclude that 12 of the 16 standard Killing spinors are also
independent of $x^-$.
The extra Killing spinors are also independent of $x^4,\dots,x^9$. This
solution has a metric which is the product of a four dimensional pp-wave
with $\bE^7$ and the flux depends on three directions in $\bE^7$.

It is interesting to observe that with the same $\zT$ we can obtain a
different
solution preserving 20 supersymmetries:
\begin{equation}
  \begin{aligned}[m]
 \zT &= n(dx^{246}+ dx^{257}  -dx^{347} +dx^{356})\\
   H=&-\tfrac{1}{9}n^2(x_2^2 +x_3^2+x_4^2+x_5^2+x_6^2+x_7^2)
-\tfrac{4}{9}n^2(x_1^2 +x_8^2+x_9^2)\\
\end{aligned}
\end{equation}
Now $\chi_4$ and $\chi_5$ are the extra Killing spinors and we have
$\lambda_4=-\lambda_5=4n$. This means that while 12 of the 16 standard
Killing spinors are 
still independent of $x^-$ the extra Killing spinors are not, and
moreover depend
on all coordinates. 

Another case peserving 20 supersymmetries with $n_1=n_2=n_3=0$ can be
obtained if we set $n_4=n_5=n_6=n_7=n$. The solution is given by
\begin{equation}
  \begin{aligned}[m]
 \zT &= n(dx^{246}+ dx^{257}  +dx^{347} +dx^{356})\\
   H=&-\tfrac{1}{9}n^2(x_1^2 +x_3^2+x_5^2+x_6^2+x_8^2+x_9^2)
-\tfrac{4}{9}n^2(x_2^2 +x_4^2+x_7^2)\\
\end{aligned}
\end{equation}
and since all $\lambda_I$ are non-zero, all Killing spinors depend on
$x^-$. The extra 4 Killing spinors depend on all $x^i$.

Finally we note that a solution with 22 supersymmetries can be found
by setting all seven of the $n_i$ to be equal to $n$. In this case
the solution is given by
\begin{equation}
  \begin{aligned}[m]
 \zT &= n(dx^{123}+dx^{145}+dx^{167} +dx^{246}+ dx^{257}  +dx^{347}
+dx^{356})\\
   H=&-n^2(x_2^2 +x_4^2+x_6^2)-\tfrac{1}{4}n^2(x_8^2+x_9^2)\\
\end{aligned}
\end{equation}
All $\lambda_I$ are non-zero and thus all Killing spinors depend on
$x^-$.
The 6 extra Killing spinors do not depend on the co-ordinates
$x_1,x_3,x_5,x_7$.

In conclusion we have demonstrated that there are solutions of
M-theory with extra supersymmetries i.e. more than 16 and less than 32.
In particular we have explicitly demonstrated solutions with
18,20,22 and 24 Killing spinors. It is possible that the
seven-parameter ansatz \p{eq:wist} allows for further possibilities, but
this seems
unlikely.
It is straightforward
to see that the Penrose limits of various intersecting branes 
with $AdS\times Sphere$ factors explicitly discussed in
\cite{Blau:2002rg} lead to special cases of our solutions.
It would be interesting to know whether all of our solutions can 
be obtained as  Penrose limits. 

Now that the forbidden
region of solutions preserving between 1/2 and all supersymmetries has
been broached here and in \cite{pope,michelson} it is natural to
wonder, as in \cite{extra}, whether all fractions are in fact
obtainable. 
Perhaps the kind of analysis of \cite{extratwo} might provide some
further 
insight into exotic fractions of supersymmetry.

{\bf Note Added}: In the final stages of writing up this work we became
aware of \cite{popecveticlu} which has significant overlap with the work 
here.



\section*{Acknowledgements}

We would like to thank Jaume Gomis for helpful discussions. 
Both authors are supported in part by PPARC
through SPG $\#$613. 


\end{document}